\documentclass[aps,prb,twocolumn,natbib,showpacs,floatfix,superscriptaddress, longbibliography]{revtex4-2}

\usepackage{graphicx}
\usepackage{amsmath,amssymb,bbold,bm,color, mathtools}
\usepackage{float}
\usepackage{epstopdf}
\usepackage{hyperref}
\usepackage{color}
\usepackage{enumerate}
\usepackage{wasysym}
\usepackage{url}
\usepackage{dsfont}
\usepackage{braket}
\usepackage{comment}
\graphicspath{{figures/}}
\hypersetup{colorlinks=true,linkcolor=blue,citecolor=blue,urlcolor=blue}

\newcommand{\up}{\uparrow}
\newcommand{\down}{\downarrow}

\newcommand{\bk}{\bm{k}}
\newcommand{\bq}{\bm{q}}

\newcommand{\bp}{\bm{p}}

\newcommand{\bQ}{\bm{Q}}

\newcommand{\cC}{{\cal C}}
\newcommand{\cT}{{\cal T}}
\newcommand{\Ctwo}{\mathcal{C}_2}

\newcommand{\cM}{{\cal M}}

\newcommand{\cH}{{\cal H}}

\newcommand{\cF}{{\cal F}}

\newcommand{\cV}{{\cal V}}
\newcommand{\cU}{{\cal U}}
\newcommand{\theM}{\theta_{\rm M}}
\newcommand{\pf}{{45^{\circ}}}

\begin{document}

\title{Twisted multilayer nodal superconductors}

\author{Tarun Tummuru}
\affiliation{Department of Physics and Astronomy \& Stewart Blusson Quantum Matter Institute, University of British Columbia, Vancouver, BC V6T 1Z4, Canada}
\affiliation{Department of Physics, University of Zurich, Winterthurerstrasse 190, Zurich 8057, Switzerland}

\author{\'Etienne Lantagne-Hurtubise}
\affiliation{Department of Physics and Astronomy \& Stewart Blusson Quantum Matter Institute, University of British Columbia, Vancouver, BC V6T 1Z4, Canada}
\affiliation{Department of Physics and Institute for Quantum Information and Matter,
California Institute of Technology, Pasadena, California 91125, USA}

\author{Marcel Franz}
\affiliation{Department of Physics and Astronomy \& Stewart Blusson Quantum Matter Institute,
University of British Columbia, Vancouver, BC V6T 1Z4, Canada}

\date{\today}

\begin{abstract}
    Twisted bilayers of nodal superconductors were recently proposed as a promising platform to host superconducting phases that spontaneously break time-reversal symmetry. Here we extend this analysis to twisted multilayers, focusing on two high-symmetry stackings with alternating ($\pm \theta$) and constant ($\theta$) twist angles. In analogy to alternating-twist multilayer graphene, the former can be mapped to twisted bilayers with renormalized interlayer couplings, along with a remnant gapless monolayer when the number of layers $L$ is odd. In contrast, the latter exhibits physics beyond twisted bilayers, including the occurrence of `magic angles' characterized by cubic band crossings when $L\mod 4 = 3$. Owing to their power-law divergent density of states, such multilayers are highly susceptible to secondary instabilities. Within a BCS mean-field theory, defined in the continuum and on a lattice, we find that both stackings host chiral topological superconductivity in extended regions of their phase diagrams.
\end{abstract}

\maketitle


\section{Introduction}

Stacking and twisting two-dimensional (2D) materials represents a new paradigm for producing a variety of emergent electronic states that are absent in the original building blocks~\cite{Balents2020, Andrei2020}. While twisted bilayer graphene is the archetypal example of this new field~\cite{Suavez2010, Bistritzer2011, Cao2018, Cao2018b}, the ideas of `twistronics' are now being applied to a wide range of van der Waals materials~\cite{Novoselov2016} including more elaborate graphene-based structures~\cite{Koshino2019, Chebrolu2019, Liu2019, Burg2019, Khalaf2019, Park2021, Hao2021}, transition metal dichalcogenides~\cite{Wu2018, Regan2020, Wang_2020, Zhang2020}, 2D magnets~\cite{Tong2018, Hejazi2020} and thin films of high-$T_c$ superconductors~\cite{Can2021, Volkov2020, Zhu2021, Zhao2021, lee2021twisted}.

Twisted bilayers of 2D nodal $d$-wave superconductors have been suggested as a candidate platform to realize topological superconductivity~\cite{Can2021, Volkov2020}, potentially extending to high temperatures comparable to the critical temperature $T_c$ of a monolayer. Within the framework of Bardeen–Cooper–Schrieffer (BCS) mean-field theory, it was shown that such a system spontaneously breaks time-reversal symmetry $\cT$ when the twist angle is close to $\pf$, and is described by a chiral order parameter of the form $d_{x^2-y^2} + i d_{xy}$ (or $d+id'$ for short). The topological gap induced in the quasiparticle spectrum may be attributed to Cooper pair co-tunneling between the rotated $d$-wave order parameters of the two layers~\cite{Yang2018, Can2021}.

Currently, the most attractive material platform to test this theory is the bismuth-based cuprate superconductor Bi$_2$Sr$_2$CaCu$_2$O$_{8+\delta}$ (Bi2212), which has been exfoliated in monolayer form with a $T_c$ close to the bulk value of $90$K~\cite{Yu2019, Zhao2019}. Experimental studies on few-layer-thick samples of twisted Bi2212 are already being reported. Transport measurements find that close to $\pf$ twist angle, the interlayer Josephson current is dominated by its second harmonic, which signals co-tunneling of Cooper pairs~\cite{Zhao2021}. While indicative of a non-trivial phase difference between the two superconducting layers, and hence $\cT$ breaking, this does not directly reveal the topological nature of the state~\cite{volkov2021, Tummuru2021}. Along similar lines, Ref.~\cite{lee2021twisted} notes a vanishing critical current at $\pf$. Another experimental work, however, interprets transport data as evidence for a subdominant $s$-wave pairing channel~\cite{Zhu2021}. Given the varying degree of agreement across these studies, additional probes such as polar Kerr angle measurements~\cite{Xia2006, Kapitulnik2009, Can2021b} and, possibly, edge current detection through local magnetometry~\cite{Wang2020} will be necessary to ascertain the nature of the superconducting phase of twisted cuprate bilayers.

On the theory side, a further complication is introduced by the well-known fact that cuprates are not quantitatively described by BCS theory. A more sophisticated treatment of correlations using a $t$-$J$ model, and the inclusion of an interlayer tunneling form factor appropriate to Bi2212, suggests a topologically trivial phase near $\pf$ that is either gapless or has a very small $\cT$-breaking gap~\cite{Song2022}. On the other hand, numerical treatment of the Hubbard model on a twisted square lattice at the commensurate angle $\theta=53.1^{\circ}$ via the variational cluster approximation finds a $\mathcal{T}$-broken phase that is gapped yet topologically trivial~\cite{Lu2021}. Discrepancies between these theoretical models highlight the importance of the complex orbital structure and strong electronic correlations inherent to cuprates.

However, it stands to reason that the phenomenology of twist-induced spontaneous $\cT$-breaking described in Refs.~\cite{Can2021, Volkov2020} is not specific to cuprates, but applies to nodal superconductors in general. Advances in exfoliation and material growth techniques could potentially reveal  new nodal superconductors in organics~\cite{Clark2010}, quasi-2D limits of iron-based~\cite{Hirschfeld2011} or heavy-fermion~\cite{White2015} materials, or in transition metal dichalcogenides following recent theoretical proposals~\cite{He2018, Wang2018, Shaffer2020}. There is thus hope that 2D nodal superconductors could form the basis for various heterostructures with novel properties~\cite{Yang2018}, possibly opening new routes towards Majorana zero-modes~\cite{mercado2022}.

In light of these developments, here we explore the physics of twisted structures comprising a generic number $L$ of nodal superconducting layers. Going beyond bilayers affords two key advantages: (i) a greater variety of possible phases with additional tunability of physical properties, and (ii) an effective renormalization of parameters that stabilizes topological superconductivity for weaker interlayer tunneling as well as smaller twist angles far from $\pf$, potentially facilitating experimental realizations.

As the twist degree of freedom of each layer can be tuned independently, multilayers admit several inequivalent arrangements. In this work we consider two high-symmetry configurations that arise by fixing a unique twist angle $\theta$ between neighboring layers: (i) alternating twists (AT), where the relative twists between successive layers differ in sign and (ii) chiral twists (CT), where neighboring layers are rotated by the same angle such that a net chirality may be associated with the system.

With the goal of shedding light on the phenomenology of twisted multilayer nodal superconductors, we forgo materials-specific considerations and theoretical complications stemming from the treatment of strong interactions, and adopt a BCS mean-field approach as a first exposition of this problem. We work mainly within a self-consistent continuum formulation, where each layer is treated as a 2D nodal superconductor with $d_{x^2-y^2}$ symmetry coupled to neighboring layers via electron tunneling. We then connect our results to a twisted multilayer lattice model, which gives us access to topological indices and edge modes. Contrasting the phase diagrams of the two stackings, we obtain the following key results.

The physics of AT multilayers can be understood via a mapping to multiple copies of the bilayer problem~\cite{Khalaf2019}. $\cT$-breaking superconductivity occurs for twist angles $\theta$ close to $\pf$, but the spectrum is fully gapped only when the number of layers $L$ is even. This is because for odd $L$ the system comprises an effectively decoupled monolayer sector with a nodal $d$-wave order parameter. In analogy to twisted bilayers, quadratic band touchings occur at a set of `magic angles'. We find that such band touchings are susceptible to $\cT$-breaking secondary instabilities through two distinct mechanisms. First, the interlayer tunneling between twisted layers tends to nucleate a $d+id'$ order~\cite{Can2021} -- an effect that is enhanced in multilayers due to a renormalization of the tunneling strength. Second, residual interactions left out of the mean-field BCS treatment of nodal superconductivity become marginal at the QBTs and can nucleate different symmetry-breaking orders~\cite{Volkov2020}.

CT multilayers spontaneously break both time-reversal $\cT$ and a $\pi$ rotation $\Ctwo$ about the diagonals of the middle layer, but preserves their product $\Ctwo \cT$. Their superconducting state is fully gapped and topological, with chiral Majorana modes propagating on sample boundaries, for generic large angles \emph{different} from $\pf$. For small twist angles, the Dirac cones in the BdG quasiparticle dispersion merge to form either quadratic or cubic band touchings, depending on $L$. The latter has an enhanced susceptibility to gap opening due a power-law divergent density of states at zero energy, as opposed to a constant density of states in the quadratic case.

The rest of this article is organized as follows. After describing the model in Sec.~\ref{sec:model}, we analyze the physics of twisted trilayers in Sec.~\ref{sec:AT} (AT stacking) and Sec.~\ref{sec:CT} (CT stacking). Sec.~\ref{sec:multilayers} presents a generalization to multilayers with $L > 3$. Finally, in Sec.~\ref{sec:outlook}, we comment on experimental signatures of the phases identified in this work and speculate on their relevance to ongoing investigations in twisted cuprates.


\begin{figure*}[t]
    \includegraphics[width=0.95\textwidth]{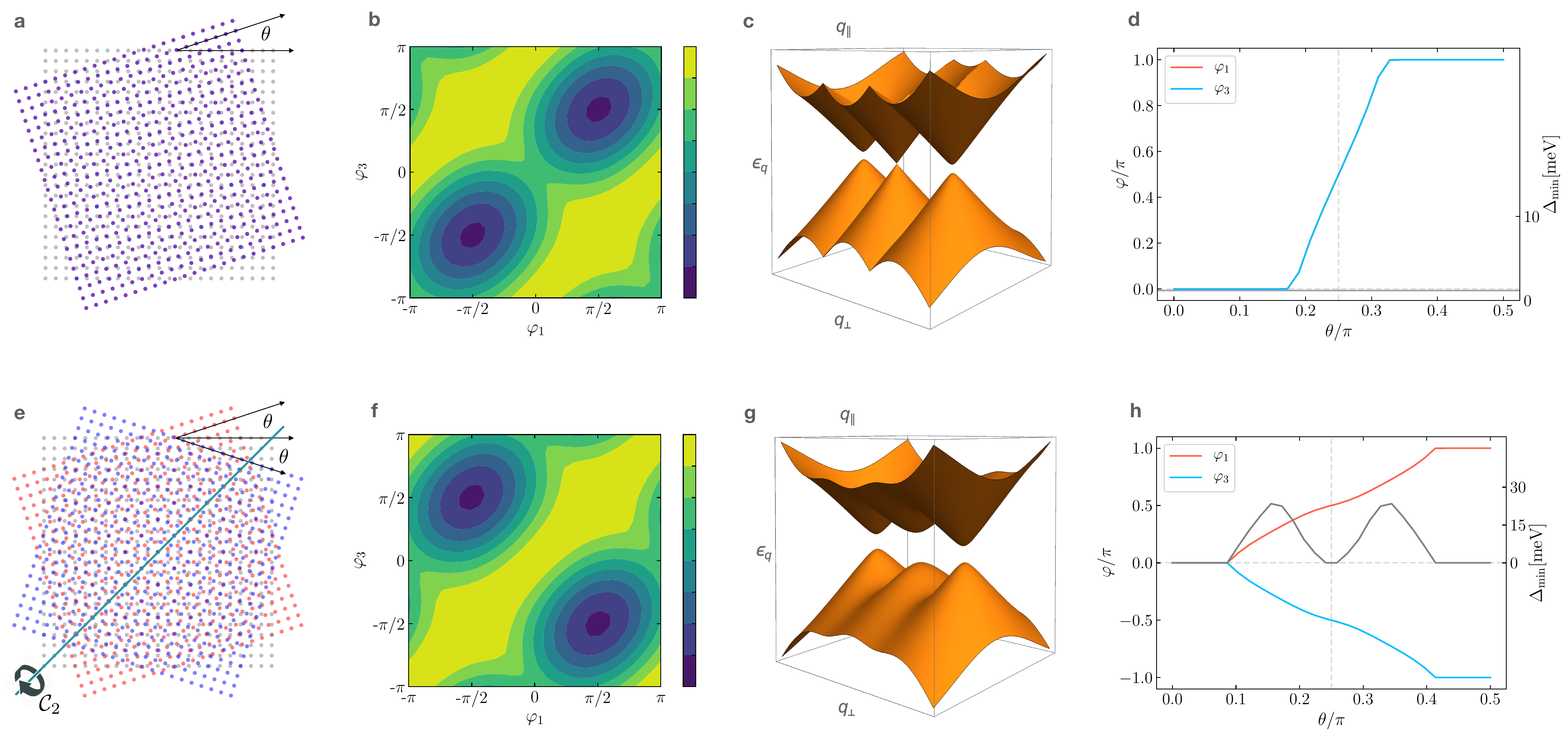}
    \caption{Spontaneous $\cT$ breaking in twisted trilayer nodal superconductors. The top panels ({\bf a}-{\bf d}) depict data for the alternating twist (AT) stacking and the bottom panels ({\bf e}-{\bf h}) for the chiral twist (CT) stacking.
    ({\bf a},{\bf e}): Lattice geometries. ({\bf a}): The AT stacking is obtained by twisting successive layers by $\pm \theta$, and is invariant under an out-of-plane mirror reflection $\cM_z$ with respect to the middle layer. ({\bf e}): The CT stacking is obtained by twisting successive layers by the same angle $\theta$ with respect to a common origin in the plane, and is invariant under the $\Ctwo$ symmetry that implements a $\pi$ rotation about the diagonals of the middle layers.
    ({\bf b},{\bf f}):
    Contour plot of the BdG free energy (in arbitrary units) for $\theta=45^{\circ}$ in the space of the phases $\varphi_1$, $\varphi_3$ of the top and bottom layer order parameters (setting $\varphi_2=0$).
    ({\bf c},{\bf g}): Low-energy spectrum in one quadrant of the BZ for large twist angles. The CT stacking is fully gapped, whereas the AT stacking is gapless due to the Dirac cone dispersion of the decoupled monolayer.
    ({\bf d},{\bf h}): Phase differences $\varphi_1$ and $\varphi_3$ that minimize the BdG free energy at zero temperature, and the spectral gap (grey curve) as a function of the twist angle. In {\bf d}, the physics is similar to that of a standalone bilayer, expect that the BdG spectrum is gapless owing to the decoupled Dirac cone. In {\bf h}, the non-trivial phases break $\cT$ but preserve the product $\Ctwo \cT$. The corresponding dispersion is gapped except at $45^{\rm o}$. The parameters are chosen as $\epsilon_c = 60$meV, $|\Delta_l| \approx 40$meV and $g=20$meV.}
    \label{fig:Tbreak}
\end{figure*}

\section{Model}
\label{sec:model}

We consider a continuum model consisting of nodal $d$-wave superconducting layers that are twisted relative to each other and coupled through single electron tunneling. In a BCS mean-field description this can be written as
\begin{align}
    \mathcal{H} &=
    \sum_{\bk \sigma l} \xi_{\bk} c^\dagger_{\bk \sigma l} c_{\bk \sigma l}
    + \sum_{\bk \sigma \langle l,m \rangle} g^{lm}_{\bk} c^\dagger_{\bk \sigma l} c_{\bk \sigma m}
    \label{eq:model_cont} \\
    & + \sum_{\bk l} \Delta_{\bk l} \left( c^\dagger_{\bk \up l} c^\dagger_{-\bk \down l} + {\rm h.c.} \right)
    - \sum_{\bk l} \Delta_{\bk l} \langle c^\dagger_{\bk \up l} c^\dagger_{-\bk \down l} \rangle. \nonumber
\end{align}
Therein, the operator $c^\dagger_{\bk \sigma l} $ creates an electron with momentum $\bk$ and spin $\sigma$ in layer $l$. For simplicity we assume a rotationally invariant Fermi surface, with the in-plane kinetic energy $\xi_{\bk} = k^2/2m_e - \mu$ (taking $\hbar=1$ throughout), and $g^{lm}_{\bk}$ denotes the electron tunneling amplitude between neighboring layers $l$ and $m$. The momentum sums span an energy interval $\xi_{\bk} \in [-\epsilon_c, \epsilon_c]$ around the Fermi level, with $\epsilon_c$ a high-energy cutoff fixed at $60$meV. We also consider a lattice model with a Fermi surface more closely resembling that of high-$T_c$ cuprate superconductors and arrive at qualitatively similar results.

The last two terms in Eq.~\eqref{eq:model_cont} originate from a mean-field BCS decoupling of an intra-layer, zero pair-momentum attractive interaction term $\cH_{\rm I} = \frac{1}{N} \sum_{\bk \bp l} V_{\bk \bp}^l c^\dagger_{\bk \up l} c^\dagger_{-\bk \down l} c_{-\bp \down l} c_{\bp \up l}$ that respects $C_4$ rotation symmetry. Here $N$ denotes the number of unit cells. We use a separable form $V_{\bk \bp}^l = - 2 \mathcal{V}  \cos(2 \alpha_{\bk} - 2 \theta_l) \cos(2 \alpha_{\bp} - 2 \theta_l)$ where $\mathcal{V} > 0$ is the interaction strength, $\theta_l$ the rigid rotation angle of layer $l$ and $\alpha_{\bk}$ is the polar angle of the wavevector $\bk$. This choice greatly simplifies our analysis;
we however expect our results to be representative of nodal superconductors more generally. This is corroborated by lattice calculations which employ a more generic, non-separable form of the interaction potential. The gap function is then defined as
\begin{align}
\Delta_{\bk l} &= \frac{1}{N} \sum_{\bp} V_{\bk \bp}^l \langle c_{-\bp \down l} c_{\bp \up l} \rangle = \Delta_l  \cos(2 \alpha_{\bk} - 2 \theta_l),
\label{eq:Delta_def}
\end{align}
where $\Delta_l \equiv  - \frac{2 \mathcal{V}}{N} \sum_{\bp} \cos(2 \alpha_{\bp} - 2 \theta_l) \langle c_{-\bp \down l} c_{\bp \up l} \rangle$ is the complex order parameter of layer $l$. The form of the tunneling matrix elements $g^{lm}_{\bk}$ depends on the symmetries of the material under consideration, and in particular of the orbitals involved in various interlayer tunneling processes~\cite{Andersen1995, Song2022}. For simplicity we focus on a momentum-independent form $g_{\bk}^{lm} = g$~\cite{Can2021,Volkov2020} and comment on its relevance for cuprate multilayers towards the end.

With these assumptions the Hamiltonian \eqref{eq:model_cont} can be written in the BdG formalism as
\begin{equation}
    \mathcal{H} = \sum_{\bk} \Psi_{\bk}^\dagger h_{\bk} \Psi_{\bk}  + E_0
    \label{eq:H_BdG}
\end{equation}
where $E_0 = \sum_{\bk l} \xi_{\bk} + \frac{N}{2 \mathcal{V}} \sum_{l} |\Delta_{l}|^2$. Trilayers are described by the Nambu spinor $\Psi_{\bk}^T = \left(c_{\bk \up 1}, c_{\bk \up 2}, c_{\bk \up 3}, c_{-\bk \down 1}^\dagger, c_{-\bk \down 2}^\dagger, c_{-\bk \down 3}^\dagger \right)$ and
\begin{equation}
	  h_{\bk}=
	 \begin{pmatrix}
	 \xi_{\bk} & g & 0 & \Delta_{\bk 1} & 0 & 0 \\
      g & \xi_{\bk} & g  & 0 & \Delta_{\bk 2} & 0 \\
      0 & g & \xi_{\bk} & 0 & 0 & \Delta_{\bk 3} \\
      \Delta^*_{\bk 1} & 0 & 0 & -\xi_{\bk} & -g & 0 \\
      0 & \Delta^*_{\bk 2} & 0 &  -g & -\xi_{\bk} & -g \\
      0 & 0 & \Delta^*_{\bk 3} & 0 & - g & -\xi_{\bk}
	  \end{pmatrix}.
\end{equation}
Analogous forms follow for multilayers with $L>3$. The order parameters $\Delta_{l}$ defined by Eq.~\ref{eq:Delta_def} are obtained self-consistently by minimizing the free energy
\begin{equation}
    \mathcal{F} = E_0 - 2 \beta^{-1} \sum_{\bk \alpha} \ln \left[2 \cosh \left(\frac{\beta E_{\bk \alpha}}{2} \right) \right],
    \label{eq:Free_Energy}
\end{equation}
where $\beta= 1/k_B T$ is the inverse temperature and $E_{\bk \alpha}$ is the positive energy band $\alpha$ obtained by diagonalizing $h_{\bk}$. Numerically, it is convenient to express the minimization conditions $\partial \mathcal{F} / \partial \Delta^*_l = 0$ as
\begin{equation}
   \Delta_l = \frac{2 \mathcal{V}}{N} \sum_{\bk \alpha} \tanh \left( \frac{\beta E_{\bk \alpha}}{2} \right) \bra{ \bk \alpha } \frac{ \partial h_{\bk} }{\partial \Delta^*_l} \ket {\bk \alpha},
   \label{eq:minimization_condition}
\end{equation}
where $E_{\bk \alpha} = \bra {\bk \alpha} h_{\bk} \ket{\bk \alpha}$ has been used to recast the minimization conditions in terms of the matrices $\partial h_{\bk} / \partial \Delta^*_l$ with $l=1 \ldots L$.

This continuum mean-field formulation is particularly attractive because it allows the twist angle to be tuned continuously, in contrast to lattice formulations in twisted geometries which in practice can only be studied at large commensurate angles (see Appendix~\ref{app:lattice_model} for details). Such lattice models are nevertheless useful to ascertain the topology of large twist angle structures by computing their Chern numbers and edge mode spectra.


\section{Trilayer: Alternating twist} \label{sec:AT}

We first consider the case of trilayers in the alternating twist configuration, where the top and bottom layers are twisted in the same direction with respect to the middle layer, as shown in Fig.~\ref{fig:Tbreak}a. This stacking is invariant under a mirror reflection symmetry $\mathcal{M}_z$ that
interchanges the top and bottom layers.

\subsection{Spontaneous $\mathcal{T}$-breaking}

In the AT geometry, the self-consistency conditions Eq.~\ref{eq:minimization_condition} yield a solution with $|\Delta_1| = |\Delta_3|$ as expected from the mirror symmetry $\cM_z$, while $|\Delta_2|$ is reduced by a factor $|\Delta_1| - |\Delta_2| \sim g^2/|\Delta_1|$. Setting the order parameter of the middle layer to be real ($\varphi_2 = 0$), we find that the preferred solution is always of the form $\varphi_1 = \varphi_3$. Much like the bilayer, non-trivial phase differences $\varphi_1 \neq 0, \pi$ that spontaneously break $\cT$ emerge for twist angles around $\theta = \pi/4$. Interestingly, however, the BdG spectrum always remains gapless.

These observations (summarized in Fig.~\ref{fig:Tbreak}a-d) can be understood by adapting results from alternating-twist multilayer graphene~\cite{Khalaf2019}. Through a unitary transformation detailed in Appendix~\ref{App:decoupling_AT}, the Hamiltonian describing the AT trilayer can be reduced to two decoupled blocks describing (i) a twisted bilayer with a renormalized interlayer coupling $\tilde{g} = \sqrt{2} g$ and (ii) a monolayer $d$-wave SC with the usual four Dirac cones in its quasiparticle spectrum. The monolayer and bilayer blocks are respectively odd and even under the mirror reflection $\cM_z$~\cite{Khalaf2019}, and thus cannot hybridize as long as $\cM_z$ is preserved. It then follows that the bilayer block admits a $\cT$-broken phase near $\theta = \pf$ via the phenomenology outlined in Ref.~\cite{Can2021}. Since the monolayer contributes spectator Dirac cones, the superconducting state however remains gapless throughout the phase diagram.

\begin{figure}[t]
	\includegraphics[width=0.85\columnwidth]{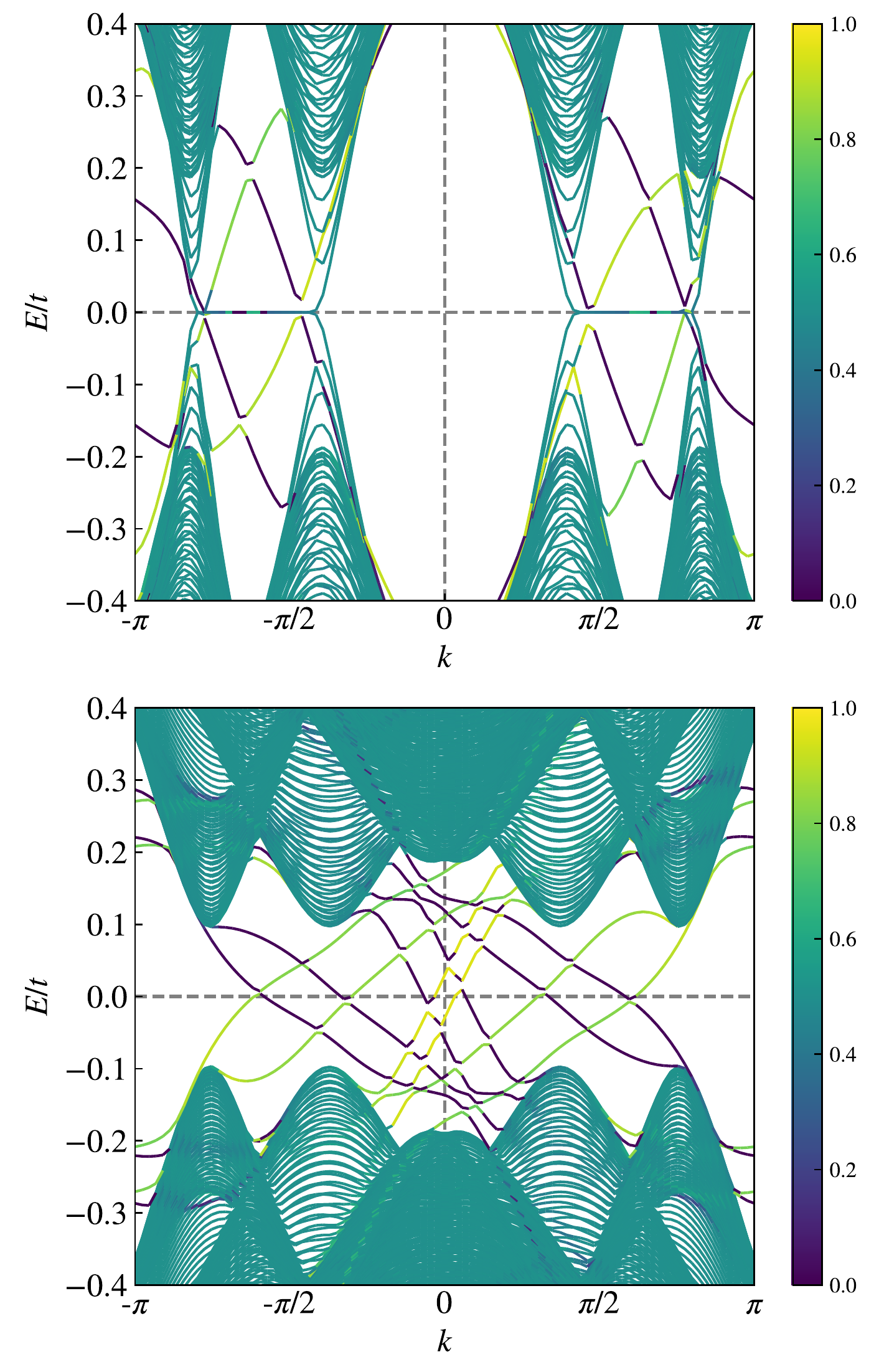}
	\caption{Spectrum of the AT trilayer (top) and the CT trilayer (bottom) on an infinite cylinder geometry for a commensurate twist angle $\theta_{1,2}=36.9^{\circ}$ (see Appendix~\ref{app:lattice_model} for modelling details). The color scale shows the expectation value $\langle \hat{y} \rangle$ of the eigenstates, with the length $y$ along the open direction normalized to unity. The AT trilayer shows $4$ chiral edge modes (contributed by the bilayer sector) that coexist with zero-energy edge states (contributed by the $d$-wave monolayer sector) connecting the projection of the bulk Dirac cones. In contrast, the CT trilayer shows a fully gapped bulk with $6$ chiral edge modes. The unit cells in the two configurations have 15 and 75 sites respectively. Both calculations use 90 unit cells along the open direction with parameters $t=1$, $\mu=-1.3$, $\Delta=0.6$ and $g=0.5$.}
	\label{fig:lattice}
\end{figure}

In the $\cT$-broken phase, the bilayer sector is characterized by a non-zero Chern number $|\cC| = 2$ or $4$, depending on the value of parameters~\cite{Can2021}. Evaluating $\cC$ using the lattice model for twisted trilayers confirms this expectation, while an infinite strip geometry shows chiral edge modes as shown in Fig.~\ref{fig:lattice}a. One subtlety in interpreting the edge mode spectra is that the monolayer sector contributes zero-energy edge modes that connect projections of the bulk Dirac cones, in analogy to zigzag graphene nanoribbons~\cite{Ryu_2002}. Chiral propagating states and zero-energy modes carry opposite eigenvalues under $\mathcal{M}_z$ and thus cannot scatter onto each other by symmetry-preserving impurities.


\subsection{Quadratic band touching}

\begin{figure*}[t]
	\centering
	\includegraphics[width=0.95\textwidth]{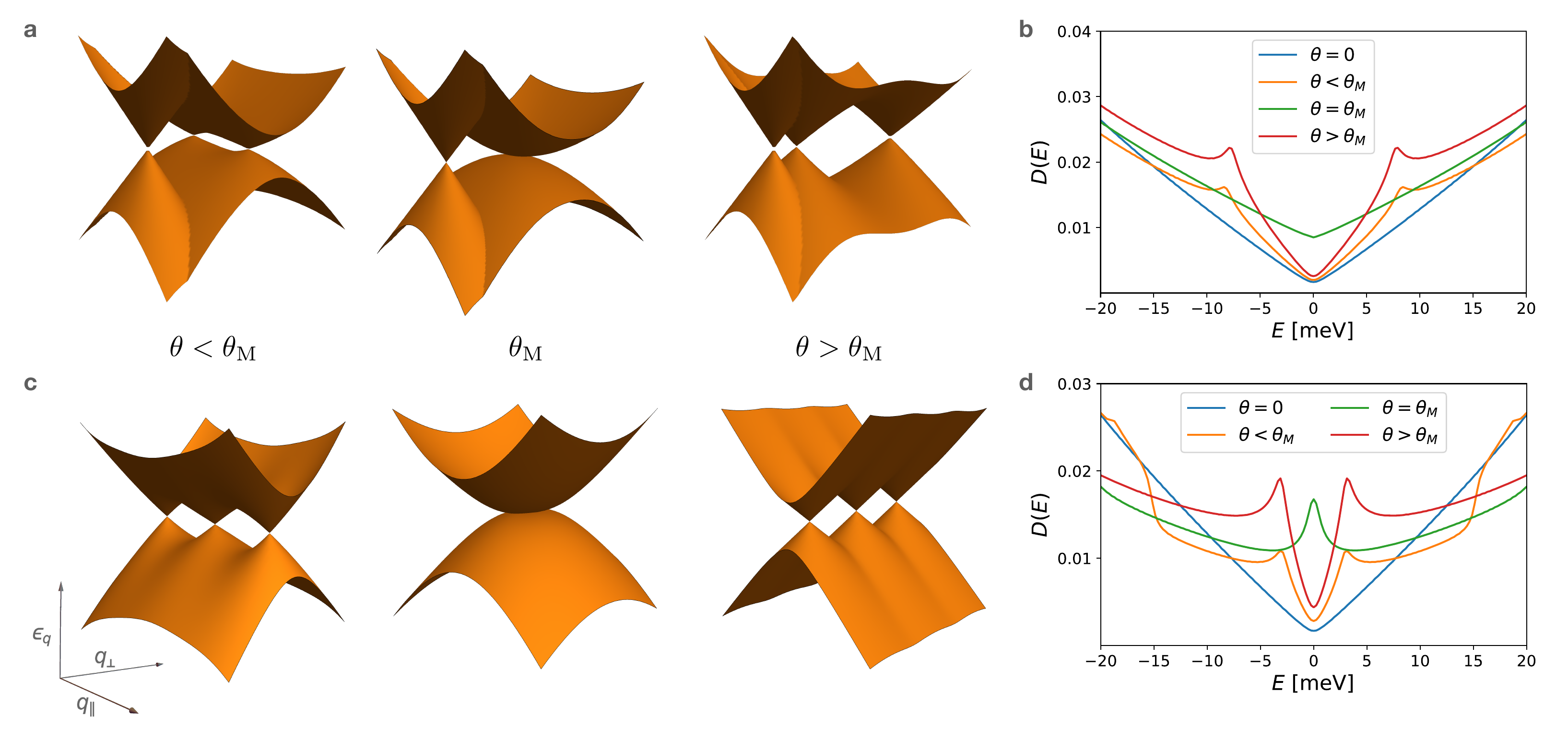}
	\caption{Dispersion of BdG quasiparticles in one quadrant of the Brillouin zone, absent any gap-opening instabilities, as the system is tuned across the magic angle $\theM$ for trilayers in the AT (top row) and CT (bottom row) configurations.
	({\bf a}):  The decoupling of AT trilayers in two sectors is manifest by the merging of two Dirac cones in a quadratic band touching (QBT) point, contributing a finite density of states at $E=0$ as shown in {\bf b}, along with a spectator Dirac cone.
	({\bf c}): The merging of three Dirac cones along the diagonals at $\bq =0$ is protected by the $\Ctwo$ symmetry of the CT stacking and leads to a cubic band touching (CBT) with a divergent density of states $D(E) \sim E^{-1/3}$ at low energies, as shown in {\bf d}.}
	\label{fig3}
\end{figure*}

The decoupled twisted bilayer sector also implies the presence of magic angles $\theM$ where Dirac cones merge pairwise to form quadratic band touchings (QBTs), as described in Ref.~\cite{Volkov2020}. To see this we linearize the Hamiltonian \eqref{eq:H_BdG} around the nodes by assuming small angles and real order parameters. In a monolayer $d$-wave superconductor, the Dirac nodes occur at the Fermi momenta $\bk_{F}$ where the order parameter changes sign. Expanding in small momenta $\bq$ around the node in the first quadrant of the BZ, the low-energy theory reads
\begin{align}
    h_{\bq} = (v_\Delta q_\perp) \tau_1 + (v_F q_\parallel) \tau_3 ,
    \label{eq:H_linearized}
\end{align}
where $\tau$ are Pauli matrices acting in Nambu space and $q_\parallel$ ($q_\perp$) denotes momentum component parallel (perpendicular) to $\bk_F$ at the node. The Fermi velocity is defined as $v_F = k_F/m_e$ and the velocity in the perpendicular direction $v_\Delta = \partial_{\alpha_{\bk}} \Delta_{\bk}/ k_F$ is evaluated at the node. The low-energy theory of a twisted multilayer is obtained by similarly linearizing all the nodes and coupling them via an interlayer tunneling matrix.

As the twist angle is tuned, two of the Dirac points in each quadrant merge into a quadratic band touching point (QBT), illustrated in Fig.~\ref{fig3}a. The magic angle is given by $\theM = 2 \tilde{g}/v_\Delta k_F$, where $\tilde{g} = \sqrt{2} g$ is the effective tunneling parameter~\cite{Volkov2020}. As outlined in Appendix~\ref{app:CBT}, the low-energy theory describing this situation can be obtained through degenerate perturbation theory and the effective Hamiltonian is given by
\begin{equation}
h^{\rm eff}_{\bq} = d_1(\bq) \eta_1 + d_3(\bq) \eta_3,
\label{eq:eff_H_QBT}
\end{equation}
where $\eta$ are Pauli matrices acting in the degenerate subspace of zero-energy solutions at the QBT, and
\begin{align}
    d_1(\bq) &= \frac{q_\parallel^2 v_{F}^2 - q_\perp^2 v_{\Delta}^2}{2 \tilde{g}} ~ , ~
    d_3(\bq) = \frac{q_\parallel q_\perp v_{F} v_{\Delta}}{\tilde{g}}.
    \label{eq:QBT_pert_theory}
\end{align}
The low-energy spectrum is quadratic and anisotropic,
\begin{align}
    E_{\bq} &= \frac{1}{2\tilde{g}}( q_\parallel^2 v_{F}^2 + q_\perp^2 v_\Delta^2 ).
\end{align}
Such two-dimensional QBTs have a non-vanishing density of states that renders them susceptible to a gap opening through spontaneous generation of a mass term~\cite{Uebelacker2011, Sun_2009}. From the perspective of the effective Hamiltonian \eqref{eq:eff_H_QBT} such a mass $M \eta_2$ breaks time-reversal symmetry and can be generated in two ways.

First is via residual interactions -- that is, interactions not taken into account by the BCS mean-field treatment of nodal superconductivity. While such sub-dominant interactions are irrelevant in the renormalization group sense for a Dirac dispersion characteristic of $d$-wave superconductors, they become marginal at a QBT. In the simplest case of an attractive $s$-wave channel with strength $\cU$, a mean-field analysis in Appendix~\ref{App:gap_opening} shows that the free energy is minimized by a $d+is$ phase, which is topologically trivial with zero Chern number although it clearly breaks $\cT$. The induced gap scales as $e^{-1/\cU D(0)}$ for small $\cU$, where $D(0) = 2 \tilde{g}/\pi v_F v_\Delta \sim \tilde{g}/\Delta$ is the density of states at zero energy at the QBT (see Appendix~\ref{App:gap_scaling}). The gap is thus exponentially small in both the interaction $\cU$ and the effective tunneling strength $\tilde{g}/\Delta$. This underlies the numerical difficultly in stabilizing this phase, unless using an unphysically large $\cU$ that is on verge of destabilizing the primary $d$-wave order~\cite{Tummuru2021}.

\begin{figure*}[t]
	\centering
	\includegraphics[width=0.95\textwidth]{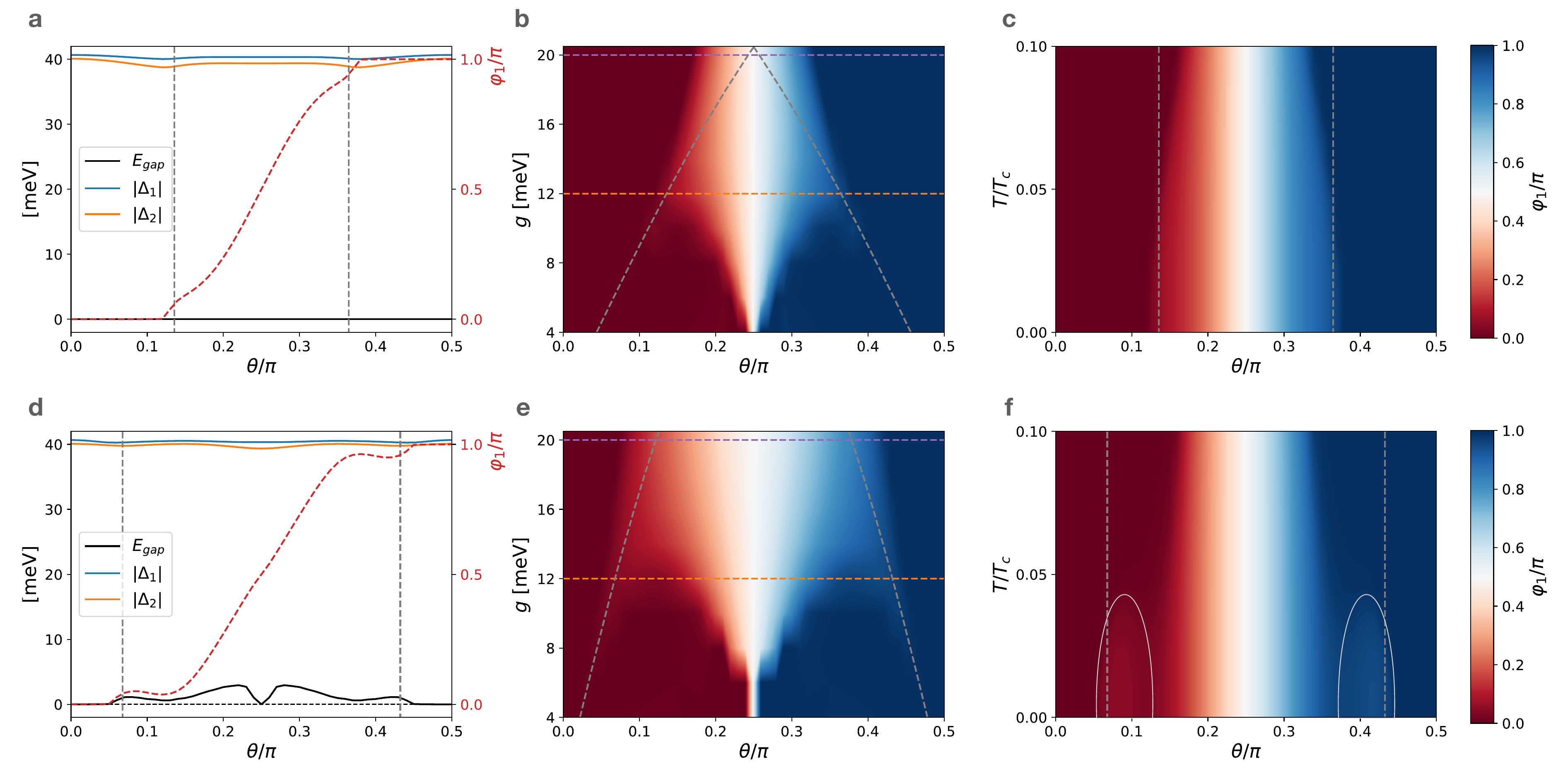}
	\caption{Self-consistent mean-field results for alternating-twist ({\bf a}-{\bf c}) and chiral-twist ({\bf d}-{\bf f}) trilayers. ({\bf a,d}): Zero-temperature phase diagrams for $g=12$meV, showing the layer-resolved order parameter amplitudes $|\Delta_l|$, phase factor $\varphi_1$ and the spectral gap $E_{\rm gap}$ as a function of $\theta$. The $\cT$-breaking phase is nucleated near the magic angles $\theM$ (vertical dotted gray lines), where the density of states is maximal, and persists to large twist angles up to $\pi/2-\theM$. For AT trilayers the spectrum remain gapless throughout due to the decoupled monolayer sector. ({\bf b,e}): Phase diagrams showing $\varphi_1/ \pi$ in the $g-\theta$ plane. The analytical magic angle conditions are indicated by dashed gray lines, and track the onset of the topological regime for small twist angles. The orange and purple lines show the $g=12$meV cut used in ({\bf c,f}) and the $g=20$meV cut used in Fig.~\ref{fig:Tbreak}, respectively. ({\bf c,f}): Phase diagrams showing $\varphi_1/\pi$ in the $T-\theta$ plane, with the temperature normalized by the critical temperature $T_c$ for zero twist angle. The $\cT$-broken domes where topological SC is stabilized around the magic angles are delineated in {\bf f}.}
	\label{fig:phase_diag}
\end{figure*}

A $\cT$-breaking gap can also open without assuming additional pairing channels beyond the leading $d_{x^2-y^2}$ channel. Indeed, Cooper pair tunneling between twisted layers generates an effective $d_{xy}$ component that can combine with the native $d_{x^2-y^2}$ order with a relative phase, via the phenomenology that underlies $\cT$-breaking at large twist angles~\cite{Can2021}. For generic small angles this mechanism is not operative because the $d_{xy}$ perturbation is irrelevant at the Dirac points. However, near the magic angles the finite density of states at the QBTs leads to the nucleation of a $d+id'$ phase. For large enough $g \gtrsim 10$meV this dome merges with that around $\pf$ to give a topological region that extends for all twist angles $\theM < \theta < \pi/2 - \theM$, as seen in Fig.~\ref{fig:phase_diag}b. This effect is enhanced in AT trilayers as compared to twisted bilayers because of the $\sqrt{2}$ increase in the effective tunneling strength, which leads to a larger magic angle and an enhanced density of states at the QBT.

Ultimately, the nature of the superconducting order parameter near the QBT will depend on the competition between the twist-angle-induced $d_{xy}$ channel and other possible sub-dominant pairing channels specific to the material under consideration.


\section{Trilayer: Chiral twist} \label{sec:CT}

\subsection{Spontaneous $\cT$-breaking}

We now consider trilayer systems with a constant twist $\theta$ between successive layers. The chiral stacking has a $\Ctwo$ symmetry corresponding to a $\pi$ rotation about diagonals of the middle layer (see Fig.~\ref{fig:Tbreak}e). In such a geometry, the problem can no longer be simplified through a unitary transformation, as in the alternating twist case, and is thus qualitatively different from twisted bilayers.

Once again choosing the order parameter of the middle layer to be real ($\varphi_2 = 0$), we find that as $\theta$ is increased the free energy develops two equivalent, time-reversed minima at $\varphi_1 = -\varphi_3 \neq 0$, as shown in Fig.~\ref{fig:Tbreak}f. This non-trivial phase structure describes a $\cT$-broken state which nevertheless preserves the product $\Ctwo \cT$. The system is gapped except at the special point $\theta = \pf$, which is gapless because the CT and AT stackings are then identical up to a rotation of the top layer by $\pi/2$, which simply contributes an additional phase difference of $\pi$ given the $C_4$ symmetry of the $d$-wave order parameter. We note that for identical parameters $g$, $\cV$ and $\epsilon_c$, the CT stacking exhibits $\cT$-breaking in a larger range of twist angles than the AT stacking, a trend which also extends to multilayers as discussed in Sec.~\ref{sec:multilayers}.

In the $\cT$ broken phase, the lattice model shows a non-zero Chern number that takes values as high as $|\cC| = 6$, depending on the choice of parameters. The corresponding chiral edge modes that traverse the bulk gap are seen in Fig.~\ref{fig:lattice}. Similar to the physics of twisted bilayers~\cite{Can2021} (whose superconducting state also respects $\Ctwo \cT$ symmetry), the Chern number assignment can be understood by noting that each layer contributes $\cC = \pm 2$ when in a $d\pm id'$ phase. Contrast this to the AT stacking where, due to presence of the $\cM_z$ mirror symmetry and the aforementioned unitary decoupling, only two out of the three layers contribute to topological superconductivity and give rise to a maximal Chern number $|\cC| = 4$.


\subsection{Cubic band touching}

We now turn to CT trilayers with small twist angles.
The $\Ctwo$ symmetry
dictates that the three Dirac cones in a BZ quadrant must be arranged symmetrically with respect to the diagonals $k_x = \pm k_y$. As illustrated in Fig.~\ref{fig3}c, at small twist angles all three Dirac cones are on the diagonals, while at large twist angles one Dirac cone remains on the diagonal and the other two are located at an equal distance on either side. The transition between these two cases occurs at the magic angle $\theM = \sqrt{2} g/ v_\Delta k_F$ where the three Dirac cones merge into a cubic band touching (CBT).

Starting from Eq.~\eqref{eq:H_linearized} we can project down to the zero-energy subspace at the CBT using degenerate perturbation theory (outlined in Appendix~\ref{app:CBT}). The resulting low-energy effective Hamiltonian reads $h^{\rm eff}_{\bq} = d_1(\bq) \eta_1 + d_3(\bq) \eta_3$ with
\begin{align}
    d_1(\bq) &= \frac{q_\perp v_{\Delta} (q_\perp^2 v_{\Delta}^2 - 3 q_\parallel^2 v_{F}^2)}{4g^2} , \nonumber \\
    d_3(\bq) &= \frac{q_\parallel v_{F} (q_\parallel^2 v_{F}^2 - 3 q_\perp^2 v_{\Delta}^2)}{4g^2}.
    \label{eq:CBT_pert_theory}
\end{align}
Note that unlike in the AT twist, here the bare interlayer tunneling $g$ appears and the spectrum takes the anisotropic cubic form
\begin{align}
    E_{\bq} &= \frac{1}{4g^2}( q_\parallel^2 v_{F}^2 + q_\perp^2 v_\Delta^2 )^{3/2}.
\end{align}

Such a CBT in two dimensions admits a divergent density of states at low energy $D(E) \sim \nu E^{-1/3}$ with $\nu = g^{4/3} / v_F v_\Delta$, as illustrated in Fig.~\ref{fig3}d, which makes the system sensitive to a gap opening. Indeed, as derived in Appendix~\ref{App:gap_scaling}, in the presence of a sub-dominant $s$-wave pairing channel of strength $\cU$ a $\cT$-breaking gap scaling as a power law $ (\cU \nu )^3 \sim \cU^3 g^4/(v_\Delta v_F)^3$ is induced. This is a stronger dependence than the exponentially small gap expected for the QBTs. However, the gap is small in practice because of the suppression by a large power of both $\cU$ and the interlayer tunneling $g$, as also demonstrated numerically in Appendix~\ref{App:gap_opening}.

In the absence of additional pairing channels, the self-consistent numerical solution shows that the system develops a $d+id'$ phase induced by interlayer tunneling for most twist angles, as shown in Fig.~\ref{fig:phase_diag}. For sufficiently large $g$ the magic angle $\theM$ marks the boundary of the $d+id'$ phase, which survives all the way to $\pi/2 - \theM$. For smaller $g$ there are two separate pockets of $d+id'$ superconductivity (one centered around $\theM$ and the other centered around $\theta = \pi/4$) that are separated by a topologically trivial ($\cT$-preserving) region. The topological superconductivity near $\theta = \pi/4$ is robust, with a critical temperature close to the native $T_c$ of a monolayer, while the domes surrounding the magic angles are more fragile and persist only up to $\sim 0.05 T_c$ for optimal $g$ and $\theta$  (see Fig.~\ref{fig:phase_diag}f).

Finally, we stress that because interactions are expected to be relevant at a CBT, a treatment of interactions beyond mean-field might be necessary to determine the fate of the system. In particular, the scale of the interaction-induced gap could be larger than that anticipated through our simple BCS treatment.


\section{Generalization to multilayers}
\label{sec:multilayers}

\begin{figure}[t]
	\centering
	\includegraphics[width=0.85\columnwidth]{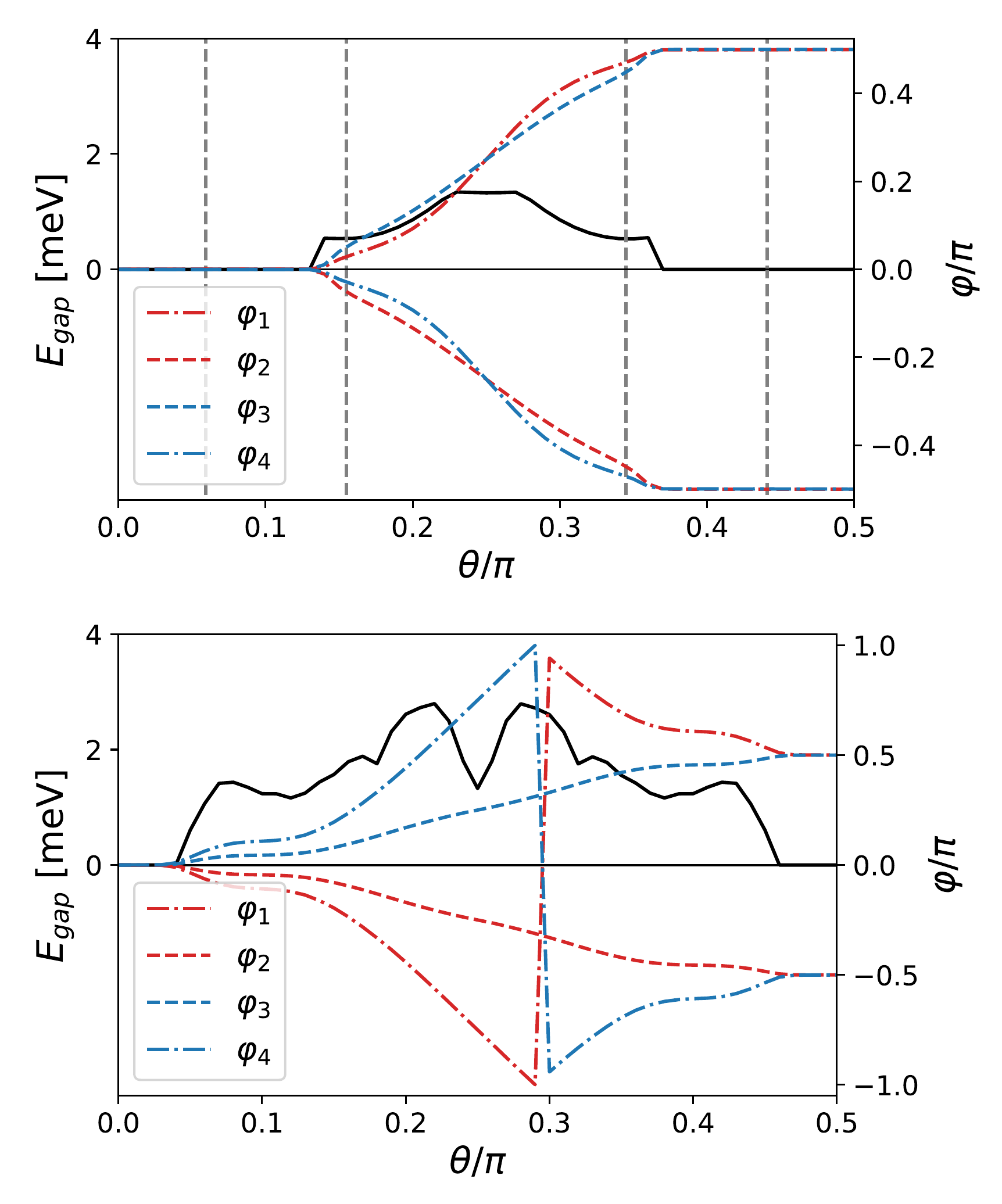}
	\caption{Self-consistent mean-field results for quadrilayers in the AT (top) and CT (bottom) configuration, using $g=12$meV, $|\Delta_l| \sim 40$meV and $\epsilon_c = 60$meV and zero temperature. For twist angles around $\theta=\pi/4$ the SC phase breaks $\cT$ for both stackings. (Top:) The magic angles are determined by the effective tunneling in the bilayer blocks $\tilde{g}_{1,2} = (\sqrt{5} \mp 1)g/2$ and are denoted by dashed gray lines. For these parameters $\cT$-breaking sets in only at the second magic angle. (Bottom:) For CT quadrilayers, chiral SC which preserves the $\Ctwo \cT$ symmetry is stabilized for a large range of twist angles.}
	\label{fig:quadrilayers}
\end{figure}

With an understanding of spontaneous $\cT$ breaking and higher-order band touchings in trilayers, we generalize our analysis to multilayer stacks.

\begin{figure*}
	\centering
	\includegraphics[width=.95\textwidth]{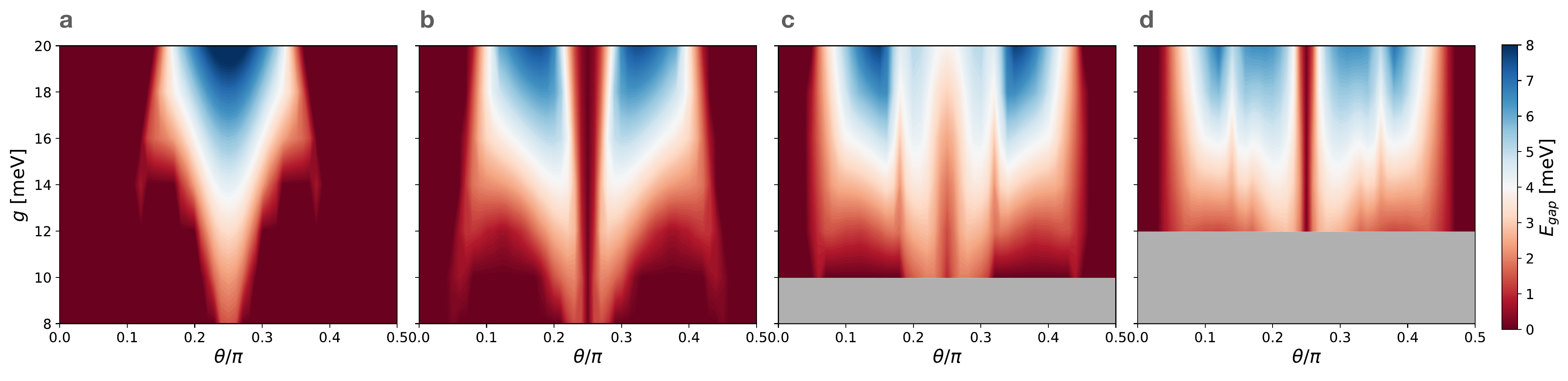}
	\caption{Phase diagram comparison of chirally twisted multilayers. The color scale indicates the $\cT$-breaking gap in ({\bf a}) bilayers, ({\bf b}) trilayers, ({\bf c}) quadrilayers and ({\bf d}) pentalayers. The extent of the topological region of the phase diagram increases with $L$, and the maximal gaps are obtained \emph{away} from $\theta = \pi/4$. We choose parameters $\epsilon_c = 60$meV, $|\Delta_l | \approx 40$meV and $T=0$. The grey areas in {\bf c} and {\bf d} denote regions where numerical convergence is difficult due to the small energy scales involved.}
	\label{fig:CT_comparison}
\end{figure*}

\subsection{Alternating twist}

In an AT stacking, layer $l$ is rotated by $\theta_l = (-1)^l \theta/2$ such that neighboring layers have relative twists $\pm \theta$. For a trilayer, the problem boiled down to a twisted bilayer and a spectator monolayer. More generally, as outlined in Appendix \ref{App:decoupling_AT}, the decoupling unitary transformation holds for any number of layers and the system reduces to $\lfloor L/2 \rfloor$ bilayers characterized by the renormalized interlayer tunneling amplitudes
\begin{equation}
    \tilde{g}_m = 2 g \cos \left( \frac{\pi m}{L+1} \right).
\end{equation}
Here $m = 1 \ldots \lfloor L/2 \rfloor$ and $\lfloor . \rfloor$ is the floor function. Additionally, in the case of odd $L$ one has a decoupled monolayer sector. When the twist angle is close to $\pf$, each of the bilayer blocks is expected to spontaneously break $\cT$, leaving any decoupled layer as is -- the layers always gap out in pairs. In the same vein, the physics at small twist angles carries over and each bilayer is characterized by a QBT at a different magic angle $\theM^m = 2 \tilde{g}_m/ v_\Delta k_F$.

The decoupling unitary transformation relies on the assumption that the order parameters in all layers of a given parity (even/odd) carry the same phase and amplitude. While such an assumption is not enforced by symmetry, it is approximately respected in the self-consistent solutions of AT multilayers, with small differences between $|\Delta_l|$ of order $g^2/\Delta$. The lack of a symmetry constraint also means that the bilayer blocks can weakly hybridize with each other, which introduces quantitative differences in the numerical values of magic angles, but does not alter the physics in a qualitative way.

An analysis of quadrilayers, summarized in Fig.~\ref{fig:quadrilayers}a, bears out these expectations. We find that the condition for unitary decoupling is approximately respected: $\Delta_1 \approx \Delta_3$ and $\Delta_2 \approx \Delta_4$. Close to $\theta=\pi/4$ a non-trivial phase difference develops between even and odd numbered layers, accompanied by a $\cT$-breaking gap. With our choice of parameters, $\cT$-breaking appears only at the second magic angle, presumably because the density of states at the first magic angle, which scales as $D(0) \sim \tilde{g}_m$, is too small to nucleate a $d+id'$ phase.
In a twisted bilayer the phase difference determines the sign of the Chern number. Given the small phase difference within the odd (even) layer suspace,
we deduce that both bilayer blocks break $\cT$ in the same way such that the total Chern number $\cC = \pm 4 \lfloor L/2 \rfloor$. This expectation is indeed borne out by our lattice calculations, which find a Chern number $|\cC| = 8$ in the $\cT$-broken phase of AT quadrilayers.

\begin{table}
\centering
\begin{tabular}{c | c | c}
  \hline
  $L$ & $N_{\rm M}$ & Low-energy spectrum at $\theM$ \\
  \hline
  1 & 0 & n/a \\
  2 & 1 &  QBT \\
  3 & 1 & CBT \\
  4 & 2 & DQBT, DQBT \\
  \hline
  5 & 2 & DQBT + Dirac cone, DQBT + Dirac cone\\
  6 & 3 & DQBT, QBT, DQBT \\
  7 & 3 & DQBT, CBT, DQBT \\
  8 & 4 & DQBT, DQBT, DQBT, DQBT \\
  \hline
\end{tabular}
\caption{In chirally twisted multilayers with weak interlayer tunneling, the low-energy spectra at the sequence of magic angles follows a mod $4$ pattern as a function of the number of layers $L$. Here $N_{\rm M}$ denotes the number of magic angles, while QBT and CBT stand for quadratic and cubic band touchings at $\bq=0$, respectively. DQBT stands for double QBT, and denotes two QBTs that occur at the same twist angle, i.e., four Dirac cones merge pairwise as shown in Fig.~\ref{fig:double_qbt}. When more than one magic angle occurs we list their corresponding low-energy spectra in order of increasing $\theM$.}
\label{tab:multilayers_mirror}
\end{table}


\begin{figure*}
	\centering
	\includegraphics[width=\textwidth]{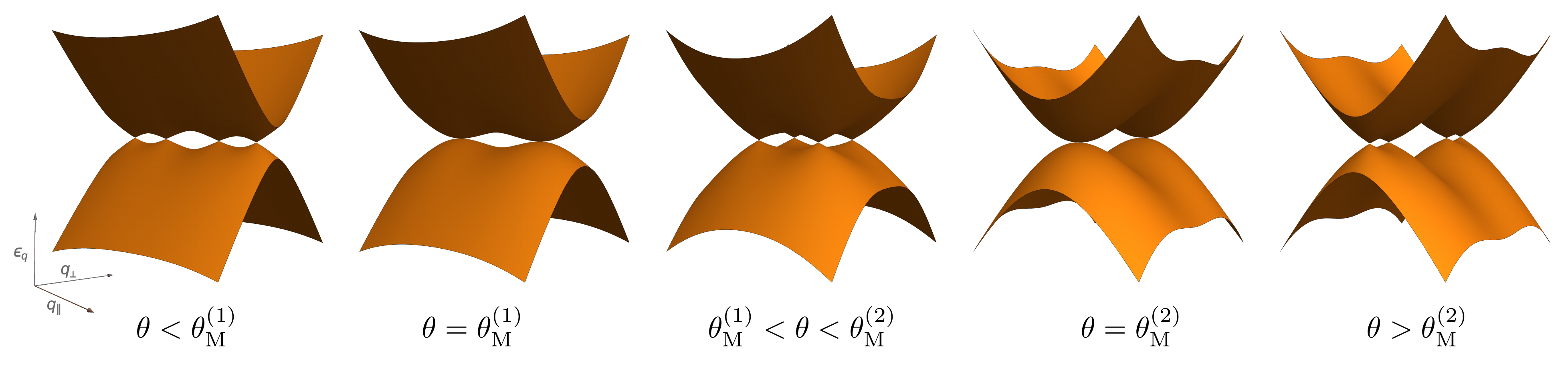}
	\caption{Evolution of the BdG quasiparticle dispersion of chirally-twisted quadrilayers at small angles. There are two magic angles $\theM^{(1)}$ and $\theM^{(2)}$ where \emph{two} quadratic band touchings occur simultaneously at momenta $\pm \bq^{(1)}_{\rm M}$ and $\pm \bq^{(2)}_{\rm M}$ respectively.}
	\label{fig:double_qbt}
\end{figure*}

\subsection{Chiral twist}

Due to the lack of a simplifying unitary transformation, the generalization to multilayers is less straightforward in the chiral twist case, which corresponds to
\begin{equation}
    \theta_l = \left(l - \frac{L+1}{2} \right) \theta.
\end{equation}
We thus explore this configuration numerically, and find that for a wide range of twist angles the system spontaneously breaks $\cT$ while respecting $\Ctwo \cT$, as depicted in Fig.~\ref{fig:quadrilayers}b for quadrilayers. Interestingly, the extent of the $\cT$-breaking region of the phase diagram increases with $L$ (see Fig.~\ref{fig:CT_comparison}). In contrast to a bilayer, the optimal twist angle for topological superconductivity occurs away from $\theta = \pi/4$. The topological gap at optimal twist angle decreases with the number of layers $L$ but the dependence is weak. Guided by the intuition that in a chirally-twisted stack all layers (except those at the very top and bottom) experience effectively the same local environment, we expect each layer to contribute the same value $\cC_l =  \pm 2$ to the total Chern number. Supported by lattice calculations on trilayers, we thus conjecture that generically $\cC = \pm 2L$ in the $\cT$-broken phase.

At small twist angles, the continuum model exhibits an intriguing pattern of band touchings linked to $L \mod 4$, summarized in Table~\ref{tab:multilayers_mirror}. We find that CBTs occur only when $L\mod 4 = 3$, and also note the occurrence of \emph{double QBTs}: two quadratic band touchings in a BZ quadrant occurring at the same twist angle. In a quadrilayer, for instance, there are two magic angles $\theM^{(1)}$ and $\theM^{(2)}$ where double QBTs arise (see Fig.~\ref{fig:double_qbt}).


\section{Conclusion and outlook}
\label{sec:outlook}

Recent theoretical works have proposed twisted nodal superconductors as a promising platform to realize topological superconducting phases. In this work we have extended these ideas to multilayers in two different geometries, alternating twist (AT) and chiral twist (CT) stackings, and mapped their phase diagrams within the framework of self-consistent BCS mean-field theory.

At small twist angles, the quasiparticle spectra of multilayers exhibit a sequence of magic angles with either quadratic, cubic or double quadratic dispersions, depending on the stacking and the number of layers $L$. The constant (power-law divergent) density of states of the quadratic (cubic) band touchings renders these systems susceptible to secondary symmetry-breaking transitions in the superconducting state, which can be triggered either by interlayer tunneling or residual interactions. While interlayer tunneling generically favors $\cT-$breaking chiral superconductivity, with an order parameter of $d+id'$ symmetry, the ordering tendencies due to the residual interactions will depend on microscopics. A full accounting of instabilities in pairing and various particle-hole channels would require a detailed understanding of interactions in a given material and might necessitate a treatment beyond mean-field.

The cubic and double quadratic band touchings constitute new features that go beyond twisted bilayers. Cubic band touchings have a divergent density of states at zero energy, $D(E) \sim E^{-1/3}$, and residual interactions are expected to be relevant from a renormalization group perspective. This is reminiscent of higher-order van Hove singularities ocurring, for instance, in the normal state dispersion of cuprates~\cite{Markiewicz2021}, Moir\'e surface states of topological insulators~\cite{Wang2021} and transition metal dichalcogenides~\cite{hsu2021}. At a double quadratic band touching, four Dirac cones in the same BZ quadrant merge pairwise to give rise to two quadratic crossings. The proximity of these nodes in momentum space could potentially lead to finite-momentum pairing instabilities.

Closer to $\theta=\pf$, both configurations spontaneously break $\cT$ but in qualitatively different ways. AT multilayers can be reduced to a set of bilayers with renormalized interlayer tunneling strengths (alongside a decoupled monolayer for odd $L$). Their physics is therefore similar to that of a twisted bilayer~\cite{Can2021, Song2022, Lu2021}. Nevertheless, in view of the estimates for the tunneling strength $g \sim 5-10$meV inferred from recent transport studies~\cite{Zhao2021, volkov2021, Tummuru2021}, such a renormalization might help stabilize $\cT$-broken superconductivity. In contrast, the problem posed by CT multilayers cannot be simplified in this way. While $\cT$ and a $\pi$ rotation $\cC_2$ along the diagonals are individually broken, their product $\Ctwo \cT$ is preserved. The system exhibits chiral topological superconductivity in an extended region of the twist-angle/interlayer tunneling phase diagram with a maximal Chern number $\cC = \pm2L$.

The connection of our results to cuprate heterostructures, explored in recent transport experiments~\cite{Zhu2021, Zhao2021, lee2021twisted}, relies on two key simplifications: the mean-field treatment of superconductivity and use of a momentum-independent interlayer tunneling form factor. The latter point was the subject of recent discussion~\cite{Song2022} in the context of twisted Bi2212~\cite{Can2021}. Close to $\pf$ twist, the fate of the SC state will depend sensitively on the symmetries of the orbitals involved in interlayer tunneling processes. In clean and aligned bilayers the direct (momentum-independent) tunneling between $d$ orbitals vanishes by symmetry at $\pf$ twist angle. There the dominant contribution is expected to come from $s$-orbital mediated tunneling, which is described by a form factor~\cite{Andersen1995} $g^{lm}_{\bk} \sim \cos(2 \alpha_{\bk} - \theta_l) \cos(2 \alpha_{\bk} - \theta_m)$ that unfortunately vanishes along the all-important nodal directions where the $\cT$-breaking gap opens~\cite{Can2021}. In the context of chirally-twisted multilayers, the vanishing of direct tunneling at $\pf$ might be less of a problem because, as illustrated in Fig.~\ref{fig:CT_comparison}, the region of the phase diagram with robust topological superconductivity is \emph{not} centered around $\pf$.

In bulk cuprates, incoherent tunneling mediated by disorder in layers interspacing CuO$_2$ planes is expected to play a crucial role in interlayer transport~\cite{Graf1993,Radtke1995,Radtke1996,Turlakov2001}. Therefore, incoherent tunneling in twisted layers is of potential interest. Because disorder breaks translation invariance, tunneling processes are not subject to symmetry constraints invoked in Ref.~\cite{Song2022} and may allow for substantial twist-induced gaps even close to $\pf$. While the full treatment of an impurity-assisted tunneling model in the twisted multilayer geometry is beyond the scope of this manuscript, a simple estimate using realistic parameters shows that a gap of order $1-10$meV could be generated in twisted bilayers~\cite{Haenel2022}.

Also, note that the analysis presented here assumes one CuO plane per layer. Current experiments, however, involve Bi2212 which has two CuO planes per monolayer. Based on the analysis in Ref.~\cite{Can2021}, we expect the phenomenology of $\cT$-breaking to hold in this situation, albeit with two quantitative modifications: (i) the scale of the gap, which is determined by a competition between inter-monolayer tunneling and intra-monolayer physics, will go down, and (ii) Chern numbers in the $\cT$-broken phases will be doubled.

In terms of experimental signatures of the topological phases predicted in this work, perhaps the most promising for cuprate heterostructures is the optical Hall conductivity probed by the polar Kerr effect, which was predicted to occur with a large magnitude owing to the gap scales involved~\cite{Can2021b}. Importantly, the Kerr effect is sensitive not only to $\cT$-breaking but also to the breaking of mirror symmetries, which allows it to distinguish between a chiral state such as $d+id'$ and a non-chiral state such as $d+is$. The predicted topological gap could additionally be detected through spectroscopies such as angle-resolved photoemission, electron tunneling, or optically using Raman response. Transport experiments such as the superconducting diode effect, whereby the critical supercurrent becomes direction-dependent, can also in principle discriminate between chiral and non-chiral $\cT$-broken states~\cite{Zinkl2021}.

Taking a broad view, the phenomenology discussed here could be relevant to other 2D nodal superconductors that can be exfoliated and stacked. The material-specific modelling of such platforms is left for future work. An interesting question concerns the creation of new heterostructures that realize an \emph{odd} number of Majorana modes, either propagating along boundaries or localized in vortex cores (see for instance Ref.~\cite{mercado2022}). The limit of a large number of layers $L$ is also of potential interest. On the one hand, similarly to alternating-twist graphene~\cite{Khalaf2019}, the `cascade' of magic angles in alternating-twist multilayers as $L$ is increased flattens the low-energy BdG bands, which could make it easier to stabilize symmetry-broken phases induced by interactions. On the other hand, the infinite-layer limit of the chiral twist stacking exhibits a non-symmorphic symmetry given by a combination of twist and translation in the stacking direction. A generalization of `3D twistronics,' developed in the context of chirally-twisted graphene multilayers~\cite{Wu2020, Cea2019, Xian2020}, to nodal superconductors thus represents an interesting open problem.


\section*{Acknowledgments}

We are grateful to O. Can, R. Haenel, P. Kim, S. Plugge and Z. Ye for illuminating discussions and correspondence. This research was supported in part by NSERC and the Canada First Research Excellence Fund, Quantum Materials and Future Technologies Program. \'E. L.-H. acknowledges support from
the Gordon and Betty Moore Foundation’s EPiQS Initiative, Grant GBMF8682.

\bibliography{ref}


\appendix


\section{Layer decoupling unitary transformation for alternating twist}
\label{App:decoupling_AT}

Here we derive a unitary transformation which, following the ideas in Ref.~\cite{Khalaf2019}, reduces alternating twist multilayers to a combination of decoupled bilayers, with an additional monolayer for odd $L$. To begin, we reorganize the layer structure in the multilayer Hamiltonian \eqref{eq:model_cont} according to an even/odd scheme
\begin{equation}
	  h_{\bk}=
	 \begin{pmatrix}
	   \xi_{\bk} \mathds{I}_{o} & W & \overline{\Delta^o_{\bk}} & 0  \\
	   W^\dagger & \xi_{\bk} \mathds{I}_{e} & 0 & \overline{\Delta^e_{\bk}} \\
	   \left( \overline{\Delta^o_{\bk}} \right)^* & 0 & -\xi_{\bk} \mathds{I}_{o} & -W \\
	   0 & \left( \overline{\Delta^e_{\bk}} \right)^* & -W^\dagger & -\xi_{\bk} \mathds{I}_{e} \\
	  \end{pmatrix}.
	\end{equation}
where $\mathds{I}_o$ ($\mathds{I}_e$) is an identity matrix of size $n_o$ ($n_e$) with $n_o$ and $n_e$ the number of layers with an odd and even index. While in the main text we take an isotropic form of $\xi_{\bk}$ for convenience, this is not necessary for the following derivation which holds as long as $\xi_{\bk}$ is layer-independent.
$W$ is a matrix of size $n_o \times n_e$ that accounts for arbitrary tunneling amplitudes between neighboring layers:
\begin{equation}
 W =  \begin{pmatrix}
      g_{12} & 0 & 0 & 0 & \ldots\\
      g_{23} & g_{34} & 0 & 0 & \ldots\\
      0 & g_{45} & g_{56} & 0 & \ldots \\
      0 & 0 & \ldots & \ldots & \ldots
    \end{pmatrix}.
\end{equation}
The superconducting order parameters have been organized into diagonal matrices in the even and odd layer subspaces as
\begin{align}
 \overline{ \Delta^o_{\bk} }& =  \text{diag}
 \begin{pmatrix}
      \Delta_{\bk, 1}, \Delta_{\bk, 3} , \ldots, \Delta_{\bk, 2 n_o - 1},
    \end{pmatrix} , \nonumber \\
 \overline{ \Delta^e_{\bk} } &=  \text{diag}
 \begin{pmatrix}
      \Delta_{\bk, 2}, \Delta_{\bk, 4} , \ldots, \Delta_{\bk, 2 n_e}.
    \end{pmatrix}
\end{align}
Given the 2D nature of superconductivity, weak interlayer tunneling is not expected to significantly alter the order parameter amplitudes across the layers. Further, in an alternating twist configuration, all layers in the odd (even) layer subspace are aligned. These observations inform the assumption that order parameters in all the odd (even) numbered layers can be treated as identical -- that is, $\overline{ \Delta^o_{\bk} } = \Delta^o_{\bk} \mathds{I}_{o}$ and $\overline{ \Delta^e_{\bk} } = \Delta^e_{\bk} \mathds{I}_{e}$. Combined with the fact that the single-particle dispersion $\xi_{\bk}$ is identical across the layers, the discussion in Ref.~\cite{Khalaf2019} inspires the following trick: transform $h_{\bk}$ to a new basis using $V = \text{diag} \left( A, B, A, B \right)$, where $A$ ($B$) is a unitary acting in the odd (even) layer subspace. We thus get
\begin{equation}
	 V^\dagger h_{\bk} V=
	 \begin{pmatrix}
        \xi_{\bk} \mathds{I}_{o} & A^\dagger W B  & \Delta^o_{\bk} \mathds{I}_{o}  & 0  \\
        B^\dagger W^\dagger A & \xi_{\bk} \mathds{I}_{e} & 0 & \Delta^e_{\bk} \mathds{I}_{e} \\
        \left( \Delta^o_{\bk} \right)^* \mathds{I}_{o}  & 0 & -\xi_{\bk} \mathds{I}_{o}  & -A^\dagger W B \\
        0 & \left( \Delta^e_{\bk} \right)^* \mathds{I}_{e} & -B^\dagger W^\dagger A & -\xi_{\bk} \mathds{I}_{e} \\
    \end{pmatrix}.
\end{equation}
The matrices $A$ and $B$ only act on the matrix of interlayer couplings $W$ because all other elements are $\sim \mathds{I}$ within the even and odd layer subspaces. We now choose $A$ and $B$ as in the singular value decomposition of $W$, with $W = A \Lambda B^\dagger$ and $\Lambda$ a $n_o \times n_e$ diagonal matrix containing the $n_e$ real eigenvalues $\lambda_m$ of $\sqrt{W^\dagger W}$. That is,
\begin{equation}
    V^\dagger h_{\bk} V=
    \begin{pmatrix}
        \xi_{\bk} \mathds{I}_{o} & \Lambda  & \Delta^o_{\bk} \mathds{I}_{o} & 0  \\
        \Lambda & \xi_{\bk} \mathds{I}_{e}  & 0 & \Delta^e_{\bk} \mathds{I}_{e} \\
        \left( \Delta^o_{\bk} \right)^* \mathds{I}_{o} & 0 & -\xi_{\bk} \mathds{I}_{o} & -\Lambda \\
        0 & \left( \Delta^e_{\bk} \right)^* \mathds{I}_{e} & -\Lambda & -\xi_{\bk} \mathds{I}_{e} \\
    \end{pmatrix}.
\end{equation}

This transformation has revealed a block structure with $n_e$ decoupled blocks describing bilayers with interlayer couplings $\lambda_m$. Note that the two `layers' in these bilayer blocks do not correspond to physical layers, but rather to a superposition of odd (even) layers determined by the $A$ ($B$) matrices in the singular value decomposition. Furthermore, when $L$ is odd ($\vert n_o - n_e \vert = 1$), the transformation additionally gives rise to a decoupled monolayer block characterized by a linear combination of the odd numbered layers.

While the above derivation holds for a generic form of the coupling matrix $W$, it becomes particularly simple when neighboring layers are coupled with identical strengths $W = g \left( \delta_{ij} + \delta_{i+1,j} \right)$. The eigenvalues of $\sqrt{W^\dagger W}$ are then given by~\cite{Khalaf2019}
\begin{equation}
    \lambda_m = \tilde{g}_m = 2 g \cos \left( \frac{\pi m}{L+1} \right)
\end{equation}
with $m = 1 \ldots n_e$. For the cases $L=3$ and $L=4$ discussed in the main text, we have $\tilde{g} = \lambda_1 = \sqrt{2} g$ and $(\tilde{g}_1, \tilde{g}_2) = (\lambda_1, \lambda_2) = \left( \frac{-1 + \sqrt{5}}{2}, \frac{1 + \sqrt{5}}{2} \right) g$, respectively.


\section{Low energy theory near higher order band crossings}
\label{app:CBT}

In order to arrive at a description of the multilayer systems near the magic angles we work in the low energy limit where the Dirac cones in the spectrum are linearized. To elucidate this first consider a monolayer $d$-wave superconductor. The Dirac cones are situated at the Fermi momentum $\bk_F$ on the BZ diagonals, where the gap function vanishes. In the first quadrant of the BZ, treating the node as the new origin and expanding in terms of small momenta $\bq$, one obtains the Dirac Hamiltonian
\begin{equation}
    h_{\bq} = (v_{\Delta} q_\perp) \tau_1 + (v_F q_\parallel) \tau_3
\end{equation}
where $\tau_j$ denote Pauli matrices acting in Nambu space and $v_F$ ($v_{\Delta}$) is the velocity in the direction parallel (perpendicular) to the BZ diagonal.


\subsection{Quadratic band touching}

To study magic angles in AT trilayers, the unitary transformation in Appendix~\ref{App:decoupling_AT} suggests that it suffices to work with a standalone bilayer with a renormalized interlayer layer coupling $\tilde{g}$.

For a moment consider an untwisted bilayer with no interlayer tunneling, i.e., $g=0$. In such a scenario, Dirac cones from the two layers overlap with each other and lie on the BZ diagonal. Focusing again on the first quadrant, we treat this nodal point as our point of reference and new origin $\bq = 0$. Now if the two layers are rotated slighty by $\pm \theta/2$, the new node locations are given by $\pm \bQ_N = (0, \pm \theta k_F/2)$ with respect to the node of an untwisted layer. Turning on the interlayer tunneling, the full low energy Hamiltonian is given by
\begin{equation}
    h =
    \begin{pmatrix}
        h_{\bq + \bQ_N} & t  \\
        t & h_{\bq - \bQ_N}
    \end{pmatrix}
\end{equation}
where the tunneling matrix $t = \tilde{g} \tau_3$. With the definition of a dimensionless parameter $\alpha = v_{\Delta} k_F \theta /\tilde{g}$, evaluating the analytical form of the dispersion is straightforward. While the full expression is uninformative, we highlight its key features: When $\alpha>2$, the two Dirac nodes are positioned symmetrically off the diagonal. For $\alpha<2$ they both lie on the diagonal. Precisely at $\alpha=2$, however, the nodes merge and give rise to a quadratic band touching (QBT). Note that our definition of $\alpha$ differs from that in Ref.~\cite{Volkov2020} by a factor of 2.

An effective theory for small momenta near the QBT can be obtained using degenerate perturbation theory. Suppose that $\ket{\alpha}$ and $\ket{\beta}$ are the degenerate zero-energy states of $h$ precisely at the node and $\ket{\mu}$ and $\ket{\nu}$ label the rest of the states. Up to third order, we then have
\begin{align}
	\bra{\beta} h^{\rm eff}_{\bq} \ket{\alpha} =&
	 \bra{\beta} h \ket{\alpha}
	+ \sum_\mu \frac{\bra{\beta} h \ket{\mu} \bra{\mu} h \ket{\alpha} }{-E_\mu} \label{eq:pert_theory} \\
	&+ \sum_{\mu, \nu}  \frac{ \bra{\beta} h \ket{\mu} \bra{\mu} h \ket{\nu} \bra{\nu} h \ket{\alpha} }{E_\mu E_\nu}. \nonumber
\end{align}
Keeping only the leading terms, which are quadratic in momentum, one arrives at the expressions \eqref{eq:eff_H_QBT} and \eqref{eq:QBT_pert_theory}.


\subsection{Cubic band touching}

For a trilayer with a chiral twist, the Dirac cones of the middle layer always lie on the BZ diagonal. Treating the cone in the first quadrant as the reference point $\bq=0$ as before, the other two nodal points are located at $\pm \bQ_N = (0, \pm \theta k_F)$. Accounting for the interlayer tunneling, the full Hamiltonian reads
\begin{equation}
    h =
    \begin{pmatrix}
        h_{\bq + \bQ_N} & t & 0 \\
        t & h_{\bq} & t  \\
        0 & t & h_{\bq - \bQ_N}
    \end{pmatrix}
\end{equation}
Unlike the bilayer, the dispersion now does not have a simple analytic form. Nevertheless, we find numerically that the Dirac cones move around as a function of $\alpha = v_\Delta k_F \theta / g$. Specifically, all three nodes lie on the normal state Fermi surface when $\alpha > \sqrt{2}$, merge into a cubic crossing at $\alpha = \sqrt{2}$ and all three fall on the diagonal when $\alpha$ is decreased further.

Using third-order perturbation theory following Eq.~\eqref{eq:pert_theory}, the spectrum can be down-folded to retain just the two low-energy bands near the cubic crossing. The low-energy effective Hamiltonian is defined by $d_1(\bq)$ and $d_3(\bq)$ in Eq.~\eqref{eq:CBT_pert_theory}.



\section{Secondary instability at higher order band crossings}
\label{App:gap_opening}

In addition to the primary $d$-wave pairing channel with strength $\cV > 0$, let us suppose that an additional pairing channel with strength $\cU > 0$ exists in each layer. We take it to be isotropic ($s$-wave) for simplicity:
\begin{equation}
H_s = - \frac{2 \mathcal{U}}{N} \sum_{\bk \bp l} c^\dagger_{\bk \up l} c^\dagger_{-\bk \down l} c_{-\bp \down l} c_{\bp \up l}.
\end{equation}
Assuming $\cU \ll \mathcal{V}$, such that the $d$-wave component remains largely unaffected, we can define the secondary $s$-wave order parameter as $\Delta^s_l = - \frac{2 \mathcal{U}}{N} \sum_{\bp} \langle c_{-\bp \down l} c_{\bp \up l} \rangle $. The mean-field Hamiltonian then reads
\begin{align}
    \cH &= E_0 + \sum_{\bk} \Psi_{\bk}^\dagger h_{\bk} \Psi_{\bk},
    \label{eq:app_BdG}
\end{align}
where $E_0 = \sum_{\bk l} \xi_{\bk} + N  \sum_{l} \left( \frac{|\Delta_{l}|^2}{2 \cV} + \frac{|\Delta^s_l|^2}{2 \cU} \right)$. Correspondingly, the full order parameter of each layer is given by
\begin{equation}
    \Delta_{\bk l} = \Delta_l \cos( 2 \alpha_{\bk} + 2 \theta_l ) + \Delta_l^s,
    \label{eq:Delta_twists}
\end{equation}
with complex $\Delta_l^s$. Recall that the free energy of the system at inverse temperature $\beta$, in the BCS mean-field approach, reads
\begin{equation}
    \mathcal{F} = E_0 - 2 \beta^{-1} \sum_{\bk \alpha} \ln \left[2 \cosh(\beta E_{\bk \alpha}/2) \right],
    \label{eq:Free_Energy_App}
\end{equation}
where $E_{\bk \alpha}$ are the positive-energy eigenvalues of the BdG Hamiltonian $h_{\bk}$ in Eq.~\ref{eq:app_BdG}. The gap equation for the secondary order parameter is obtained by minimizing $\cF$ with respect to $\Delta_l^{s *}$:
\begin{equation}
   \Delta_l^s = \frac{2 \cU}{N} \sum_{\bk \alpha} \tanh \left( \frac{\beta E_{\bk \alpha} }{2} \right) \bra{ \bk \alpha } \frac{ \partial h_{\bk} }{\partial \Delta_l^{s *}} \ket {\bk \alpha} .
\end{equation}

\begin{figure}
	\centering
	\includegraphics[width=0.85\columnwidth]{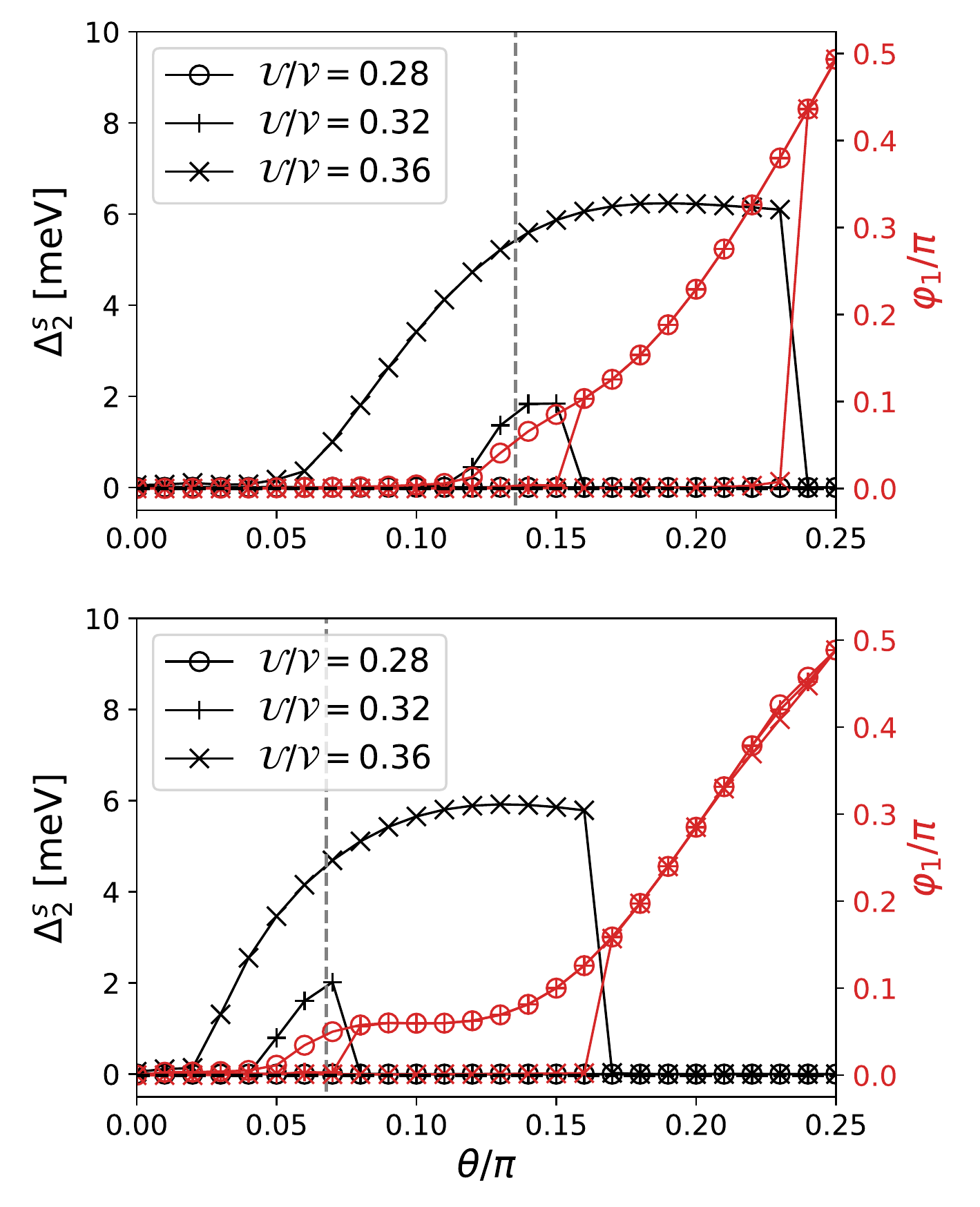}
	\caption{Self-consistent mean-field results for the $s$-wave order parameter amplitude $\Delta^s_2$ in the middle layer (black lines) and the $d$-wave phase factor $\varphi_1$ (red lines) as a function of twist angle $\theta$ for trilayers in the AT (top panel) and CT (bottom panel) stacking. We consider a secondary $s$-wave pairing channel with relative strength $\cU/\cV = 0.28$, $0.32$ and $0.36$, and use parameters $g=12$meV and $\epsilon_c=60$meV, while the primary $d$-wave order parameter $\Delta \sim 40$meV. For $\cU/\cV \lesssim 0.30$ we find that the $d+is$ solution is not nucleated, and the $d+id'$ solution first develops around the magic angles. For large values of $\cU$ we observe a competition between the two gapped $\cT$-broken phases, with the $d+id'$ solution preferred at large angles and the $d+is$ solution preferred near the magic angles.}
	\label{fig:trilayers_swave}
\end{figure}

In Fig.~\ref{fig:trilayers_swave} we show phase diagrams for trilayers in both the AT and CT stackings, in the presence of the secondary instability. In our numerics we simultaneously solve for the $s$-wave and $d$-wave components, $\Delta_l^s$ and $\Delta_l$. We find that $\Delta^s$ acquires a purely imaginary value that results in a gapped spectrum, for both QBTs (as obtained in AT trilayers) and CBTs (as obtained in CT trilayers) at their respective magic angles. Crucially, we find that the $\cT$ breaking secondary instability is nucleated only for large $\cU/\cV \sim 0.3$ in our numerics, which can be understood in terms of an exponential (power-law) suppression of the induced gap for quadratic (cubic) band touchings, as we clarify in the following Appendix.

\section{Density of states and gap scaling at higher order band crossings}
\label{App:gap_scaling}

Working in the low-energy limit, it is possible to derive simple analytical scaling forms for the gap opening induced by residual interactions at the quadratic and cubic band touchings. Since self-consistent numerics from Appendix~\ref{App:gap_opening} indicate that the $\cT$-breaking $s$-wave order parameter is purely imaginary, the effective low energy Hamiltonian may be written as
\begin{equation}
    \cH^{\rm eff} = \frac{N}{2\cU}M^2 + \sum_{\bq} [d_1(\bq) \eta_1 + M \eta_2 + d_3(\bq) \eta_3],
\end{equation}
where the $\bq$ dependence of $d_1(\bq)$ and $d_3(\bq)$ is determined by the type of crossing as described in the main text and Appendix~\ref{app:CBT}. Assuming that the solution in the $d$-wave channel remains unchanged, one can minimize the effective free energy with respect to the mass term as
\begin{equation}
    \frac{\partial \cF^{\rm eff}}{\partial M} = N \frac{M}{\cU} - \sum_{\bq} \left[  \tanh \left( \frac{\beta E_{\bq}}{2} \right) \frac{\partial E_{\bq}}{\partial M} \right] = 0,
\end{equation}
and thus
\begin{equation}
    M = \cU \int \frac{d^2 \bq}{(2 \pi)^2} \tanh \left( \frac{\beta E_{\bq}}{2} \right) \frac{\partial E_{\bq}}{\partial M},
\end{equation}
where we transformed $\frac{1}{N} \sum_{\bq} \rightarrow \int d^2 \bq/(2 \pi)^2$. We then express the integral in energy domain using $\int d^2 \bq/(2\pi)^2 \rightarrow \int dE ~ D(E)$ with the density of states $D(E)$, and consider the zero-temperature limit ($\beta \to \infty$) where
\begin{equation}
    M = \cU \int dE ~ D(E) \frac{\partial E}{\partial M}.
\end{equation}
Let us now focus on the two cases of interest.


\subsection{Quadratic band touching}

The QBT, described at low energies by the effective Hamiltonian Eq.~\eqref{eq:QBT_pert_theory}, consists of an anisotropic dispersion $E_{\bq} = \frac{1}{2\tilde{g}} ( q_\parallel^2 v_F^2 + q_\perp^2 v_\Delta^2 )$. A generic cross section of this dispersion is an ellipse, with major and minor axes given by $a=\sqrt{2 \tilde{g}E}/v_F$ and $b= \sqrt{2 \tilde{g}E}/v_\Delta$ respectively. The density of states is easily obtained by a standard trick: let us consider the quasiparticle density at energy $E$ defined as $\rho(E) = \int_0^E dE' D(E')$, which is simply given by the area enclosed by the ellipse at energy $E$ normalized by the momentum spacing,
\begin{equation}
    \rho(E) = \pi a b  \left( \frac{1}{2 \pi} \right)^2 = \frac{\tilde{g} E}{2 \pi v_F v_\Delta}.
\end{equation}
Differentiating and taking into account the $4$ QBTs in the Brillouin zone, we find the total DOS
\begin{equation}
    D(E) = \frac{2 \tilde{g}}{\pi v_F v_\Delta}.
\end{equation}
Using the derivative of the gapped spectrum $E_{\bq} = \sqrt{d_1(\bq)^2 + d_3(\bq)^2 + M^2}$ with respect to the induced mass term $\partial{E} / \partial M = M / E$, and noting that the integration over energy ranges from the band bottom to a high-energy cutoff $\Lambda \gg M$, this leads to
\begin{equation}
    \frac{\pi v_F v_\Delta}{2 \cU \tilde{g}} = \ln \Lambda - \ln M,
\end{equation}
which is just the BCS scaling form
\begin{equation}
    M = \Lambda \text{exp} \left( {-\frac{\pi v_F v_\Delta}{2 \cU \tilde{g}} } \right).
\end{equation}
This derivation shows that the induced gap $M$ is exponentially small in the product of the attractive strength $\cU$ and the effective interlayer tunneling $\tilde{g}$.


\subsection{Cubic band touching}

The CBT is described at low energies by the effective Hamiltonian Eq.~\eqref{eq:CBT_pert_theory} with anisotropic dispersion $E_{\bq} = \frac{1}{4 g^2} ( q_\parallel^2 v_F^2 + q_\perp^2 v_\Delta^2 )^{3/2}$. The cross section is again an ellipse, but the major and minor axes scale differently with energy, $a= (4 g^2 E)^{1/3}/v_F$ and $b= (4 g^2 E)^{1/3}/v_\Delta$ respectively. The number of states enclosed by the ellipse is
\begin{equation}
    \rho(E) = \frac{ab}{4 \pi}= \frac{ (4g^2 E)^{2/3}}{4 \pi v_F v_\Delta},
\end{equation}
which leads to the power-law divergent density of states (again counting the four QBTs in the Brillouin zone)
\begin{equation}
    D(E) = \frac{2 (2 g)^{4/3} }{3 \pi v_F v_\Delta}  E^{-1/3}.
\end{equation}
And since $\partial{E} / \partial M = M/ E$, we have
\begin{align}
    \frac{3 \pi v_F v_\Delta}{2 (2 g)^{4/3} \cU} &= \int_M^{\Lambda} dE \frac{1}{E^{4/3}}.
\end{align}
The above integral is UV convergent, and evaluates to $3 M^{-1/3}$ in the limit $\Lambda/M \rightarrow \infty$. We thus find
\begin{equation}
    M = \frac{(2 g)^{4} (2 \cU)^3}{\left(\pi v_F v_\Delta\right)^3},
\end{equation}
a power-law dependence of the $\cT$-breaking order parameter on the $s$-wave interaction strength $\cU$. This power-law dependence is asymptotically stronger than the exponential scaling for the QBT, but is nevertheless strongly supressed at small $g$ due to the scaling $M \sim g^4$.


\section{Lattice model}
\label{app:lattice_model}

For the case of trilayer systems we further use a twisted lattice model at commensurate twist angles to corroborate the findings of our continuum model analysis. This approach allows us to compute the Chern number $\mathcal{C}$ of the $\cT$ broken phases, as well as probe the edge states on an infinite cylinder geometry. Following the approach outlined in Ref.~\cite{Can2021} we consider a twisted square lattice Hubbard model with nearest neighbor density-density electron interactions. After a mean-field decoupling, one obtains a $d$-wave superconductor in each layer,
\begin{align}
  \cH  = & -t \sum_{\langle ij \rangle \sigma l} c^\dag_{i \sigma l} c_{j \sigma l} - t'\sum_{\langle \langle ij \rangle \rangle \sigma l} c^\dag_{i \sigma l} c_{j \sigma l}
  - \mu \sum_{i \sigma l} n_{i \sigma l} \nonumber \\
  &+ \sum_{\langle ij \rangle l}\left(\Delta_{ij, l}c^\dag_{i\uparrow
      l}c^\dag_{j\downarrow l}+{\rm h.c.}\right)
- \sum_{i j \sigma \langle l m \rangle} g_{ij} c^\dag_{i \sigma l} c_{j \sigma m},
    \label{hm_latt1}
\end{align}
where $l$ is a layer index, $t$ and $t'$ are the nearest-neighbor and next-nearest-neighbor hopping amplitudes on the square lattice, $\mu$ is the chemical potential that controls on-site particle density $n_{i \sigma l}$ and $\Delta_{ij,l}$ the complex order parameter on the bond connecting sites $i$ and $j$. For the numerical results presented in the main text we set $t'=0$ for simplicity, which leads to a single electron-like Fermi surface centered around $\bk=0$ for each monolayer. We also checked that our results are not qualitatively altered by modeling instead the Fermi surface of near-optimally-doped Bi2212, which comprises four hole pockets centered around wavevectors $(\pm \pi, \pm \pi)$, by setting the parameters $t' = -0.45t$ and $\mu = -1.35t$ following Ref.~\cite{Bille2001}. For interlayer tunneling, we assume an exponentially decaying form $g_{ij} = e^{-(r_{ij}-c)/\rho}$ that is circularly symmetric and connects sites separated by $r_{ij}$; the scale is set by parameters $c=2.2$ and $\rho=0.4$ in units of the lattice constant~\cite{Can2021}.

To make use of the Bloch representation of wavefunctions in the lattice model, one needs to work with commensurate twist angles such that the multilayer system forms a crystallographic unit cell. For alternating twists, the commensurability condition, which is same as that for the bilayer, can be specified in terms of a twist vector $(m, n)$ with integers $m,n$. The corresponding twist angle is $\theta_{m,n} = 2 \arctan(m/n)$ and the number of sites in the unit cell is given by $L(m^2+n^2)$.

In the chiral twist case, a periodic lattice forms between neighboring layers for the same $\theta_{m,n}$ derived above; however, there is now the additional constraint that the non-adjacent layers must also form a supercell that is commensurate. Let us consider a trilayer and denote the twist between layers 1-2 and 1-3 by vectors $(m,n)$ and $(p,q)$, respectively. We are interested in the case where $\theta_{p,q} = 2 \theta_{m,n}$, that is,
\begin{align}
	\arctan({p/ q}) = 2 \arctan({m/ n}),
	\label{eq:comm_cond}
\end{align}
which, using the trigonometric expansion for $\tan(2\theta)$, simplifies to
\begin{align}
	\frac{p}{q} = \frac{2(m/n)}{1-(m/n)^2}.
\end{align}
This implies that the supercell corresponding to the $(m,n)$ chirally twist trilayer comprises $3(p^2+q^2)$ lattice sites. When $(m,n)=(1,2)$, which corresponds to a twist angle $\theta_{1,2} = 36.9$ as in Fig.~\ref{fig:lattice}, $(p,q) = (4,3)$ and the smallest unit cell has 75 sites.

This exercise can be extended to $L=4$, for instance, by solving for the relations $\theta_{p,q} = 2 \theta_{m,n}$ and $\theta_{r,s} = 3 \theta_{m,n}$, where $(r,s)$ now denotes the twist vector for layers 1-4.


\end{document}